\documentclass{iopart}

\usepackage{graphicx}
\usepackage{array}
\usepackage{iopams} 
\usepackage{bm}
\usepackage[english]{babel}
\usepackage[latin1]{inputenc}

\begin{document}

\title[]{Auxiliary field method for the square root potential}

\author{Bernard Silvestre-Brac$^1$, Claude Semay$^2$ and Fabien Buisseret$^2$}

\address{$^1$ LPSC Universit\'{e} Joseph Fourier, Grenoble 1,
CNRS/IN2P3, Institut Polytechnique de Grenoble, 
Avenue des Martyrs 53, F-38026 Grenoble-Cedex, France}
\address{$^2$ Groupe de Physique Nucl\'{e}aire Th\'{e}orique, Universit\'{e}
de Mons, Acad\'{e}mie universitaire Wallonie-Bruxelles, Place du Parc 20,
B-7000 Mons, Belgium}
\eads{\mailto{silvestre@lpsc.in2p3.fr}, \mailto{claude.semay@umh.ac.be}, 
\mailto{fabien.buisseret@umh.ac.be}} 

\begin{abstract}
Using the auxiliary field method, we give an analytical expression for the
eigenenergies of a system composed of two non-relativistic particles interacting via a potential of
type $\sqrt{a^2 r^2 + b}$. This situation is
usual in the case of hybrid mesons in which the quark-antiquark pair evolves in
an excited gluonic field. Asymptotic expressions are proposed and the approximate
results are compared to the exact ones. It is shown that the accuracy is excellent.
\end{abstract}

\submitto{J. Phys. A: Math. Theor.}
\maketitle

\section{Introduction}

The auxiliary field method (AFM) has been recently developed and used to compute
analytical approximate relations for bound state eigenvalues of the Schr\"{o}dinger
equation. Formulas very accurate are obtained for power-law potentials \cite{Sil08a}, 
sums of two power-law potentials \cite{Sil08b} and exponential potentials \cite{Sil08c}.
More recently it has been shown that, although obtained in very different ways, the AFM method 
and the envelope theory \cite{Hal83} are completely equivalent \cite{Bui08}. Let us note that
this equivalence leads to a deeper understanding of both frameworks. 

The aim of this report is to give an analytical expression of the eigenenergies of 
the Schr\"{o}dinger equation with the potential 
\begin{equation}
\label{eq:potex}
V(r)=\sqrt{a^2r^2+b},
\end{equation}
using the AFM. Such an interaction has a strong interest in hadronic physics, in
particular for hybrid mesons in which the quark-antiquark pair evolves in
an excited gluonic field. The applications of the results obtained here for
potential~(\ref{eq:potex}) to such systems will be given elsewhere \cite{Sem08}. 

\section{Eigenenergies}
\label{sec:Eigen}

\subsection{Analytical expression}
\label{subsec:analexp}

Let us follow the general procedure of the AFM \cite{Sil08a}. 
Our goal is to find approximate expressions
for the eigenvalues of the Hamiltonian
\begin{equation}
\label{eq:H}
H=\frac{\bm{p}^2}{2m} + V(r),
\end{equation}
where $m$ is the reduced mass of the particles and $V(r)$ is given by (\ref{eq:potex}).
We first choose an auxiliary function $P(r)=r^2$; the auxiliary field $\nu$ is then defined by
\begin{equation}
\label{eq:funcK}
\nu = K(r) = \frac{V'(r)}{P'(r)} = \frac{a^2}{2 \sqrt{a^2r^2+b}}.
\end{equation}
For the moment $\nu$ is an operator, and (\ref{eq:funcK}) can be inverted to give
$r$ as a function of $\nu$: $r = I(\nu)$. Explicitly
\begin{equation}
\label{eq:funcI}
I(\nu)=\sqrt{\frac{a^2}{4 \nu^2}-\frac{b}{a^2}}.
\end{equation}

The AFM needs the definition of a Hamiltonian $\tilde{H}(\nu) = \bm{p}^2/(2m) + \nu P(r) +
V(I(\nu))-\nu P(I(\nu))$. In our particular case,
\begin{equation}
\label{eq:tildeH}
\tilde{H}(\nu) = \frac{\bm{p}^2}{2m} + \nu r^2 + \frac{a^2}{4 \nu} + \frac{b \nu}{a^2}.
\end{equation}
If we choose the auxiliary field in order to extremize $\tilde{H}$: $\delta \tilde{H}/
\delta \nu |_{\nu = \hat{\nu}} = 0$, then the value of this Hamiltonian for such an
extremum is precisely the original Hamiltonian: $\tilde{H}(\hat{\nu}) = H$. Instead of
considering the auxiliary field as an operator, let us consider it as a real number.
In this case, the eigenenergies of $\tilde{H}$ are exactly known for all $(n,l)$ quantum
numbers:
\begin{equation}
\label{eq:eigenener}
E(\nu)=\sqrt{\frac{2 N^2 \nu}{m}}+ \frac{a^2}{4 \nu} + \frac{b \nu}{a^2},
\end{equation}
where, as usual, $N=2n+l+3/2$ is the principal quantum number of the state.

The philosophy of the AFM is very similar to a mean field  procedure. We first seek the value
$\nu_0$ of the auxiliary field which minimizes the energy, $\partial E/\partial \nu
|_{\nu=\nu_0}$, and consider that the value $E(\nu_0)$ is a good approximation
of the exact eigenvalue.
It is useful to use the new variable
\begin{equation}
\label{eq:defx0}
x_0 = a^{2/3} \left( \frac{m}{2N^2}\right)^{1/6} \nu_0^{-1/2}
\end{equation}
and to define the parameter
\begin{equation}
\label{eq:defY}
Y=\frac{16b}{3}\left(\frac{m}{2a^2N^2}\right)^{2/3}.
\end{equation}
The minimization condition is concerned now with the $x_0$ quantity and
results from the fourth order reduced equation
\begin{equation}
\label{eq:redeq}
4 x_0^4 - 8 x_0 - 3Y = 0.
\end{equation}
The solution of this equation can be obtained by standard algebraic techniques.
It looks like
\begin{equation}
\label{eq:valx0}
x_0 = G(Y) =  \frac{1}{2} \sqrt{V(Y)} + \frac{1}{2} \sqrt{ 4 (V(Y))^{-1/2}
- V(Y)},
\end{equation}
with
\begin{equation}
\label{eq:defVY}
V(Y)=\left(2 + \sqrt{4 + Y^3} \right)^{1/3} - Y \left(2 + \sqrt{4 + Y^3}
\right)^{-1/3}.
\end{equation}
Substituting this value into the expression of $E(\nu_0)$ leads to the analytical
form of the searched eigenenergies, namely
\begin{eqnarray}
\label{eq:Enu0}
E(\nu_0)&=&\frac{1}{2 G(Y)^2} \sqrt{\frac{b}{3Y}} \left[12 G(Y) + 3 Y \right] \nonumber \\
&=&2\sqrt{\frac{b}{3 Y}} \left[G^2(Y)+\frac{1}{G(Y)} \right].
\end{eqnarray}
The problem is entirely solved.

As it is shown in \cite{Sil08b}, the same formula would be obtained for the choice 
$P(r)=\textrm{sgn} (\lambda)r^\lambda$ with $\lambda > -2$, but with different 
forms for the quantity $N$. With the choice $\lambda = 2$ made above, $N=2 n+l+3/2$. 
In this case, using results from \cite{Bui08}, it can be shown that formula~(\ref{eq:Enu0}) 
gives an upper bound of the exact result. For $\lambda = -1$, $N=n+l+1$ and 
the formula gives a lower bound. The qualities of these bounds are examined below. 
The form of $N$ is not exactly analytically known for other values of $\lambda$.

An approximate simpler form of (\ref{eq:Enu0}) avoiding the complicated $G$ function and  
giving the lowest order exact results in both limits 
$Y\to 0$ and $Y\to \infty$ for finite value of $b$, is given by 
\begin{equation}
\label{eq:Enu0alt}
E(\nu_0)\approx \sqrt{\frac{b}{3Y}} \left( \sqrt{3 Y+(3\times2^{2/3}-\eta)^2}+\eta \right),
\end{equation}
where $\eta$ is an arbitrary parameter.
A very good approximation is obtained for $\eta$ around 1: for a fixed value of $b$, the relative error 
between (\ref{eq:Enu0}) and (\ref{eq:Enu0alt}) is below 2\%.

\subsection{Asymptotic expansions}
\label{subsec:asymp}

At long range, the potential (\ref{eq:potex}) behaves as the linear
potential $ar$. This asymptotic behavior is equivalent to the limit $b \to 0$.
In this case $Y \to 0$ but $b/(3Y) \to (2a^2N^2/m)^{2/3}/16$.
Moreover $G(Y) \to 2^{1/3}$ when $Y \to 0$. Reporting these conditions
in the value $E(\nu_0)$ given by (\ref{eq:Enu0}), one obtains the asymptotic
behavior
\begin{equation}
\label{eq:asymbehav1}
E(\nu_0)^2 \approx \frac{9}{4} \left(\frac{a^2}{m}\right)^{2/3} N^{4/3}.
\end{equation}
This is precisely what is expected for a pure linear potential $ar$ (see i.e.
\cite{Sil08a}). Very accurate estimation of the exact eigenvalues can then be obtained 
by using, for instance, $N=(\pi/\sqrt{3}) n+l+\sqrt{3}\pi/4$ \cite{Sil08a,Sil08b}.

Another interesting asymptotic expression is the limit $b \to \infty$, or equivalently
$a \to 0$. It is easy to check that, in this limit, the potential (\ref{eq:potex})
reduces to
\begin{equation}
\label{eq:potasym}
V(r) \approx \sqrt{b} + \frac{a^2 r^2}{2 \sqrt{b}}.
\end{equation}
Thus the potential is just equivalent to a harmonic oscillator plus a constant term.
Consequently, the exact energies are given by
\begin{equation}
\label{eq:exactoscl}
E^*=\frac{a}{\sqrt{m \sqrt{b}}} \left( 2 n+l+\frac{3}{2} \right) + \sqrt{b}.
\end{equation}
When $a \to 0$, then $Y \to \infty$ and it is easy to see that
$G(Y) \approx (3Y/4)^{1/4}$. For this limit, (\ref{eq:Enu0}) reduces to (\ref{eq:exactoscl})
for the choice $N=2 n+l+3/2$.
In this case also, the AFM leads to the good asymptotic behavior.
 
\section{Scaling laws}
\label{sec:scal}

Let us denote by $E^*(m,a,b)$ the exact eigenvalue of the Hamiltonian~(\ref{eq:H}). 
The scaling properties of the Schr\"{o}dinger equation (see
\cite{Sil08b}) allow to express it in term of the eigenenergy of a reduced equation.
More precisely
\begin{equation}
\label{eq:emab}
E^*(m,a,b) = \left( \frac{2a^2}{m}\right)^{1/3} \epsilon^*\left( b
\left( \frac{m}{2a^2}\right)^{2/3} \right),
\end{equation}
where $\epsilon^*(\beta)$ is the exact eigenvalue of the dimensionless Hamiltonian
\begin{equation}
\label{eq:Hbeta}
H(\beta)=\frac{\bm{q}^2}{4} + \sqrt{x^2+\beta}.
\end{equation}

Switching to the AFM approximation, one can verify that the approximate
energy $E(m,a,b)$ as given by (\ref{eq:Enu0}) satisfies the same scaling law
(\ref{eq:emab}) as the exact energy, the reduced approximate energy
$\epsilon(\beta)$ being still given by (\ref{eq:Enu0}) in which 
$E=\epsilon$, $b=\beta$ and
\begin{equation}
\label{eq:simplY}
Y= \frac{16 \beta}{3 N^{4/3}}.
\end{equation}
In consequence, to test the quality of the approximation it is sufficient
to make comparisons between $\epsilon(\beta)$ and $\epsilon^*(\beta)$.
This is the subject of the next section.

The Fourier transform of the Hamiltonian $H=\frac{\bm{p}^2}{2m} +\sqrt{a^2 r^2+b}$ is a 
spinless Salpeter Hamiltonian with a harmonic potential
\begin{equation}
\label{eq:sS}
H_{\textrm{S}}=\omega\sqrt{\bm{p}^2+M^2}+\sigma r^2,
\end{equation}
where the following substitutions have been made \cite{Li05}
\begin{equation}
\label{eq:sust}
\omega =\left( \frac{4 a}{m^2} \right)^{1/3}, \quad M= \frac{\sqrt{b}}{\omega}, \quad \sigma = \frac{m a \omega}{8}.
\end{equation}
In order to get a relevant equation for particles of mass $M$, $\omega$ must be set to 1 or 2. 
So the results obtained here can also be used to study the spectra of the 
Hamiltonian~(\ref{eq:sS}). Such a task will be developed in another work where the AFM
will be applied to relativistic Hamiltonians \cite{Sil09}.

\section{Comparison to exact results}
\label{sec:comp}

As remarked previously, the AFM cannot give strong constraints on the dependence
of $N$ in terms of $(n,l)$. In particular, had we chosen $P(r)=\textrm{sgn}(\lambda)\,
r^\lambda$, the better choice for $N$ would have been $N = A(\lambda)n + l +
C(\lambda)$, with the quantities $A(\lambda)$ and $C(\lambda)$ given in
\cite{Sil08a}. The square root potential ensures a smooth transition from a
linear form ($\lambda=1$ but in this case we have only approximate expressions) to
a quadratic form ($\lambda=2$ and in this case the values are exact) as $\beta$
increases from $0$ to $\infty$.

It is thus natural to suppose a smooth dependence of these coefficients on the
only relevant parameter of the problem, namely $\beta$. Therefore we choose, for
$N$ appearing in $Y$ through (\ref{eq:simplY}), an expression in the form
\begin{equation}
\label{eq:exprN}
N=A(\beta) n + l + C(\beta).
\end{equation}
From the results of \cite{Sil08a}, it is expected that 
$A(0)\approx \pi/\sqrt{3}\approx 1.814$, 
$C(0)\approx \sqrt{3}\pi/4\approx 1.360$, 
$\lim_{\beta\to\infty}A(\beta)=2$ and 
$\lim_{\beta\to\infty}C(\beta)=3/2$.

\begin{figure}[ht]
\begin{center}
\includegraphics*[width=6.4cm]{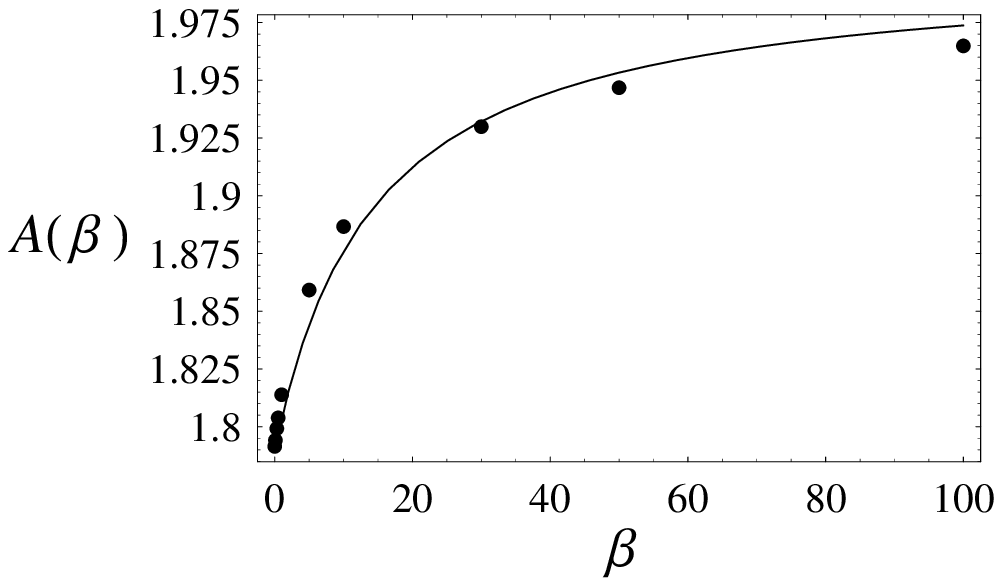}
\includegraphics*[width=6.4cm]{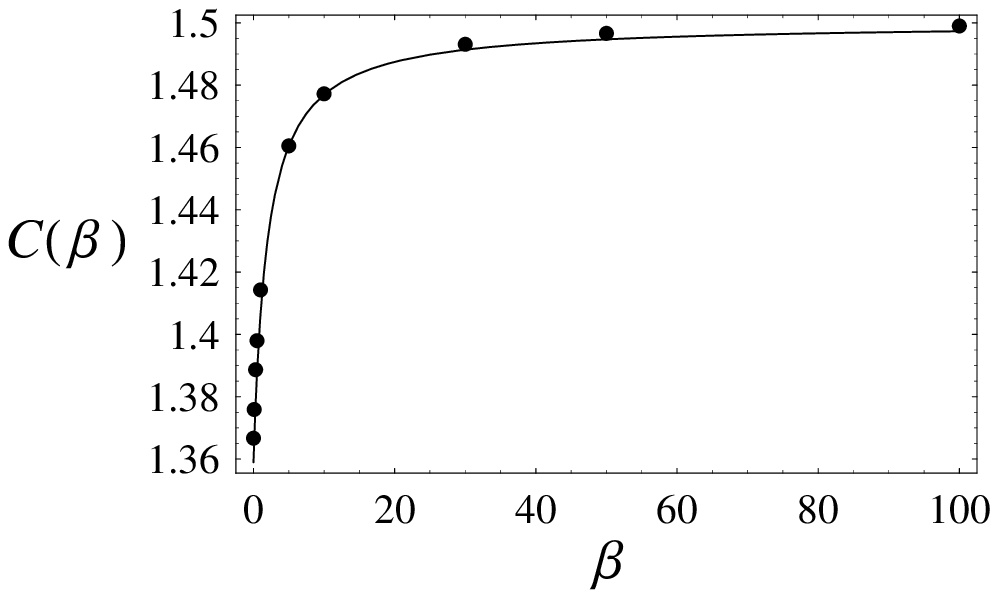}
\caption{\label{fig:exp} Best values of the coefficients $A(\beta)$ and $C(\beta)$
to parameterize the eigenvalues of Hamiltonian~(\ref{eq:Hbeta}): 
numerical fit with (\ref{eq:chis}) (dots); functions (\ref{eq:coefacd}) (solid line).} 
\end{center}
\end{figure}

The procedure we adopt is based on the following points:
\begin{itemize}
  \item We calculate the exact values $\epsilon_{nl}^*(\beta)$ for $0 \leq n \leq
n_{\max}$, $0 \leq l \leq l_{\max}$ and for a given set of $\beta$ values. This
program is achieved using a very powerful method known as the Lagrange mesh
method (described in detail in \cite{Sem01}). For our purpose, we consider
that $n_{\max}=l_{\max}=4$ is a good choice. For any calculated value, we have
an accuracy better than $10^{-5}$.
  \item We calculate the approximate values $\epsilon_{nl}(\beta)$ using (\ref{eq:Enu0})
with $Y$ given by (\ref{eq:simplY}) in which $N$ is deduced from (\ref{eq:exprN}),
for the same set of $\beta$ values. Building the $\chi$-square
\begin{equation}
\label{eq:chis}
\chi(\beta) = \frac{1}{(n_{\max}+1)(l_{\max}+1)}\sum_{n=0}^{n_{\max}} \sum_{l=0}^{l_{\max}}
\left(\epsilon_{nl}^*(\beta)-\epsilon_{nl}(\beta)\right)^2,
\end{equation}
we request the coefficients $A(\beta)$ and $C(\beta)$ of $N$ to minimize this function. 
The obtained values are represented by black dots in figures~\ref{fig:exp}.
  \item In order to obtain functions which are as simple as possible, continuous in $\beta$,
and which reproduce at best the above calculated values, we choose hyperbolic forms
and require a best fit on the set of the sample. Explicitly, we find
\begin{equation}
\label{eq:coefacd}
A(\beta) = \frac{8 \beta + 102}{4 \beta + 57}, \quad
C(\beta) = \frac{30 \beta + 53}{20 \beta + 39}.
\end{equation}
These integers are rounded numbers whose magnitude is chosen in order to not exceed too much 100.
The corresponding values are plotted as continuous curves in figures~\ref{fig:exp}.
They have been constrained to exhibit the right behavior $A \to 2$ and
$C \to 3/2$ for very large values of $\beta$. Formulas (\ref{eq:coefacd}) give $A(0)=102/57\approx 1.789$ and
$C(0)=53/39\approx 1.359$. Both values are very close to the theoretical numbers given above.
\end{itemize}

\begin{figure}[ht]
\begin{center}
\includegraphics*[width=6.4cm]{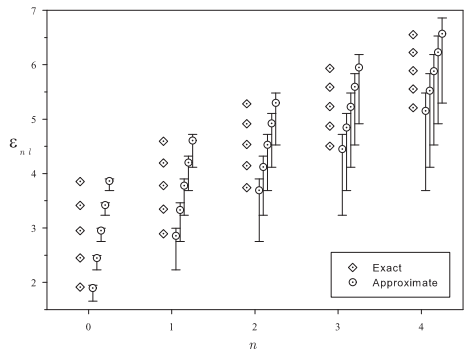}
\caption{\label{fig:comp} Spectra $\epsilon_{nl}$ of Hamiltonian~(\ref{eq:Hbeta}) with $\beta=1$.
For each value of $n$, the eigenvalues are presented for $l$ varying from 0 to 4. 
Diamonds: exact results obtained by numerical integration.
Circles: approximate results given by (\ref{eq:Enu0}) with (\ref{eq:simplY}), (\ref{eq:exprN}) and
(\ref{eq:coefacd}). The error bars extend from lower bounds obtained with $N=n+l+1$ to 
upper bounds obtained with $N=2 n+l+3/2$.} 
\end{center}
\end{figure}

Since our results are exact for $\beta \to \infty$, one has obviously $\chi=0$
in this limit. The error is maximal for small values of $\beta$ but, over the
whole range of $\beta$ values, the results given by 
our analytical expression can be considered as excellent.
Just to exhibit a quantitative comparison, we report in table~\ref{tab:comp} and 
in figure~\ref{fig:comp}
the exact $\epsilon_{nl}^*(\beta)$ and approximate $\epsilon_{nl}(\beta)$ values
obtained for $\beta=1$, a value for which the corresponding potential is neither
linear nor harmonic. 
As can be seen, our approximate expressions are better than $1\%$ for any
value of $n$ and $l$ quantum numbers. Such a good description is
general and valid whatever the parameter $\beta$ chosen.

\begin{table}[ht]
\caption{\label{tab:comp} Comparison between the exact values $\epsilon_{nl}^*(\beta)$ (2nd line)
and analytical approximate expressions $\epsilon_{nl}(\beta)$ for the
eigenvalue of Hamiltonian~(\ref{eq:Hbeta}) with $\beta=1$. For each set $(n,l)$, the exact result
is obtained by numerical integration. 3rd line: approximate results
are given by (\ref{eq:Enu0}) with (\ref{eq:simplY}), (\ref{eq:exprN}) and
(\ref{eq:coefacd}); 1st line: upper bounds obtained with $N=2 n+l+3/2$; 4th line:
lower bounds obtained with $N=n+l+1$. }
\begin{indented}
\item[]\begin{tabular}{@{}cccccc}
\br
    & $l=0$ & $l=1$ & $l=2)$ & $l=3$ & $l=4$\\
\mr
$n=0$ & 1.94926 & 2.49495 & 2.99541 & 3.46197 & 3.90193 \\
      & 1.91247 & 2.45074 & 2.94841 & 3.41419 & 3.85430 \\
      & 1.89549 & 2.44621 & 2.95032 & 3.41969 & 3.86189 \\
      & 1.65395 & 2.22870 & 2.75000 & 3.23240 & 3.68492 \\

$n=1$ & 2.99541 & 3.46197 & 3.90193 & 4.32027 & 4.72059 \\
      & 2.89556 & 3.34652 & 3.77899 & 4.19405 & 4.59335 \\
      & 2.85420 & 3.32970 & 3.77678 & 4.20097 & 4.60620 \\
      & 2.22870 & 2.75000 & 3.23240 & 3.68492 & 4.11355 \\

$n=2$ & 3.90193 & 4.32027 & 4.72059 & 5.10556 & 5.47723 \\
      & 3.74112 & 4.14232 & 4.53310 & 4.91307 & 5.28251 \\
      & 3.69078 & 4.11913 & 4.52783 & 4.91998 & 5.29790 \\
      & 2.75000 & 3.23240 & 3.68492 & 4.11355 & 4.52250 \\

$n=3$ & 4.72059 & 5.10556 & 5.47723 & 5.83725 & 6.18692 \\
      & 4.50374 & 4.87138 & 5.23246 & 5.58628 & 5.93264 \\
      & 4.44883 & 4.84403 & 5.22459 & 5.59242 & 5.94903 \\
      & 3.23240 & 3.68492 & 4.11355 & 4.52250 & 4.91485 \\

$n=4$ & 5.47723 & 5.83725 & 6.18692 & 6.52732 & 6.85935 \\
      & 5.20859 & 5.55148 & 5.88996 & 6.22329 & 6.55111 \\
      & 5.15078 & 5.52098 & 5.87970 & 6.22821 & 6.56756 \\
      & 3.68492 & 4.11355 & 4.52250 & 4.91485 & 5.29295 \\
\br
\end{tabular}
\end{indented}
\end{table}

The upper bounds obtained with $P(r)=r^2$ are far better than the lower bounds computed with $P(r)=-1/r$.
This is expected since the potential $\sqrt{a^2 r^2+b}$ is closer to a harmonic interaction than to a Coulomb one.
Better lower bounds could be obtained with $P(r)=r$. But, 
the exact form of $N$ is not known for this potential, except for $l=0$ for which $N$ can be 
expressed in term of zeros of the Airy function. With the approximate form 
$N=(\pi/\sqrt{3}) n+l+\sqrt{3}\pi/4$ \cite{Sil08a,Sil08b}, we have checked that results obtained are good but 
the variational character cannot be guaranteed.

\section{Conclusions}

In this paper, we propose an analytical approximate expression for the
eigenenergies of a Schr\"{o}dinger equation for two non-relativistic particles
interacting via a potential of type $\sqrt{a^2 r^2 + b}$. This situation corresponds
to the case of a hybrid meson in which the quark-antiquark pair evolves in an excited
gluonic field \cite{Sem08}. We give the corresponding expressions for any value of the
parameters $a$ and $b$ and for any values of the radial $n$ and orbital $l$
quantum numbers. Thanks to a Fourier transform, the energy spectrum we find can also
describe a relativistic one-body or two-body Hamiltonian with a harmonic potential.

The scaling laws properties are shown to be fulfilled exactly by these approximate
expressions; moreover the limiting cases $b \to 0$ (linear potential) and $b \to
\infty$ (harmonic potential) reduce to the exact solutions.

The approximate analytical results are compared to the exact ones. It is shown
that for any values of the parameters and for a whole range of quantum numbers
the obtained accuracy is excellent. The formulas we get are expected to play an
important role in the identification of hybrid mesons \cite{Sem08}.

\ack

CS and FB thank the F.R.S.-FNRS for financial support.


\section*{References}

\begin{thebibliography}{99}

\bibitem{Sil08a} Silvestre-Brac B, Semay C and Buisseret F 2008
\textit{J. Phys. A} \textbf{41} 275301
\bibitem{Sil08b} Silvestre-Brac B, Semay C and Buisseret F 2008
\textit{J. Phys. A} \textbf{41} 425301
\bibitem{Sil08c} Silvestre-Brac B, Semay C and Buisseret F 2008 Auxiliary field method and analytical solutions of the Schrödinger equation with exponential potentials (arXiv:0811.0287)
\bibitem{Hal83} Hall R L 1983 \textit{J. Math. Phys.} \textbf{24} 324;
1984 \textit{J. Math. Phys.} \textbf{25} 2708
\bibitem{Bui08} Buisseret F, Semay C and Silvestre-Brac B 2008 Equivalence between the auxiliary field method and the envelope theory (arXiv:0811.0748)
\bibitem{Sem08} Semay C, Buisseret F and Silvestre-Brac B 2008 Towers of hybrid mesons (arXiv:0812.3291)
\bibitem{Li05} Li Z-F, Liu J-J, Lucha W, Ma W-G and Schoberl F F 2005 \textit{J. Math. Phys.} \textbf{46} 103514
\bibitem{Sil09} Silvestre-Brac B, Semay C and Buisseret 2009 Analytical eigenvalues of semirelativistic Hamiltonians with the auxiliary field method \textit{in preparation} 
\bibitem{Sem01} Semay C, Baye D, Hesse M and Silvestre-Brac B 2001
\textit{Phys. Rev. E} \textbf{64} 016703 

\end{thebibliography}
\end{document}